\documentclass[12pt,preprint]{aastex}

\usepackage{CJK}
\begin{document}
\begin{CJK}{UTF8}{gbsn}

\title {Evidence for an  Anhydrous Carbonaceous  Extrasolar Minor Planet }

\author{M. Jura\altaffilmark{a}, P. Dufour\altaffilmark{b},   S. Xu\altaffilmark{a,c}(许\CJKfamily{bsmi}偲\CJKfamily{gbsn}艺),  B. Zuckerman\altaffilmark{a}, B. Klein\altaffilmark{a}, E. D. Young\altaffilmark{d}, \& C. Melis\altaffilmark{e}
}

\altaffiltext{a}{Department of Physics and Astronomy, University of California, Los Angeles CA 90095-1562; jura@astro.ucla.edu, kleinb@astro.ucla.edu, ben@astro.ucla.edu}
\altaffiltext{b}{D\'{e}partement de Physique, Universit\'{e} de  Montr\'{e}al, Montr\'{e}al, Qu\'{e}bec H3C 3J7, Canada; dufourpa@astro.umontreal.ca}
\altaffiltext{c}{European Southern Observatory (ESO), Garching, Germany; sxu@eso.org}
\altaffiltext{d}{Department of Earth, Planetary, and Space Sciences, University of California, Los Angeles, Los Angeles CA 90095, eyoung@ess.ucla.edu}
\altaffiltext{e}{Center for Astrophysics and Space Sciences, University of California, San Diego, CA 92093-0424; cmelis@ucsd.edu}

\begin{abstract}
Using Keck/HIRES, we report abundances of 11 different elements heavier than helium in the spectrum of Ton 345, a white dwarf that has accreted one of its own minor planets.  This particular extrasolar planetesimal which was at least 60\% as massive as Vesta  appears to have been carbon-rich and water-poor; we suggest it was compositionally similar to  those Kuiper Belt Objects with relatively little ice. 
  \end{abstract}
\section{INTRODUCTION}

 Because of the vagaries of gravitational dynamics that occur within  a white dwarf's  planetary system, a minor planet's orbit can be strongly perturbed so that it passes  close enough to the host star to be tidally disrupted \citep{Debes2002,Bonsor2011,Veras2012,Frewen2014}.  A circumstellar disk is then formed, and the white dwarf ultimately accretes the resulting debris, thereby imparting a 
 signature in the stellar spectrum which would otherwise be essentially pure hydrogen or, less often, pure helium \citep{Jura2003}.  By determining the abundances of heavy elements in the atmosphere of the host white dwarf, we can measure
the bulk elemental compositions of extrasolar minor planets to investigate their history and evolution and to compare and contrast with solar system bodies \citep{Jura2014}.

Carbon and oxygen are the most abundant heavy elements in planet forming environments.  However, they  can be carried in volatile molecules and are not automatically incorporated into rocky planetesimals.   As a result, many
observational and theoretical efforts have been devoted toward understanding the fate of  these elements in protoplanetary disks \citep{Bergin2013,Henning2013}.  
In the simplest standard picture, most oxygen is combined into water and is retained onto solids exterior to the snow line.   In these same environments, if interstellar  solid carbon grains are destroyed as occurred within the inner solar system \citep{Lee2010}, most carbon is carried in volatile molecules such as CH$_{4}$ and CO$_{2}$.   Consequently, little  carbon  accumulates into planetesimals unless they form in an extremely cold environment far from the central star.    Therefore, within asteroids,  carbon is expected to be significantly more depleted than oxygen, as typically found in the inner solar system  \citep{Lee2010} and extrasolar planetesimals  \citep{Jura2014,Xu2014a}.

This simple snow line scenario for carbon/oxygen ratios is not universally valid.  Anhydrous  Interplanetary Dust Particles  (IDPs), among the most
primitive material in the solar system,   are relatively carbon-rich and are thought to derive from  comets despite their lack of hydrous minerals  \citep{Thomas1993}.   These anhydrous IDPs might therefore be related to those Kuiper Belt Objects  (KBOs) such as
 Haumea \citep{Lacerda2007,Lockwood2014} and Eris \citep{Sicardy2011} that are  sufficiently dense to be no more than 15\%  ice by mass \citep{Brown2012} even though they likely contain large amounts of
 carbon.  Here, we suggest that the minor planet being accreted onto Ton 345 is compositionally similar to  an ice-poor KBO.

 \section{TON 345}
  Originally identified as a faint blue star at high galactic latitude, Ton 345 ( = WD 0842+231) with m(g) = 15.73 mag in the Sloan Digital Sky Survey (SDSS) has an atmosphere composed almost entirely of helium.    Ton 345  was singled out to be of special interest because it displays broad emission lines characteristic of a circumstellar
gaseous  disk orbiting within the tidal radius of the central white dwarf \citep{Gaensicke2008}.   Subsequently, this star also has been found to have excess infrared emission produced
  by an orbiting dust disk \citep{Brinkworth2012, Farihi2010, Melis2010}.   Experience has shown that at high spectral resolution, multiple elements can be detected in white dwarfs  that display excess infrared emission, and we
  therefore obtained  spectra at the Keck I telescope. 
    
  As listed by \citet{Xu2013}, there are four well-studied white dwarfs with helium dominated atmospheres, dust disks, and more
  than 10 elements heavier than He detected in their atmospheres: GD 362, GD 40, PG 1225-079 and WD J0738+1835, the only one of these four also to display a gaseous component to its circumstellar disk.  Here, we present results for Ton 345, an additional white dwarf with these distinctive characteristics.

 \section{OBSERVATIONS}
 
The data reported here were acquired in 2008 at the Keck I telescope with HIRES \citep{Vogt1994}, an echelle spectrograph with a spectral resolution near 40,000.  Table 4 of \citet{Melis2010} 
provides the exact exposure times and dates.  In total, we obtained 6600 s and 9000 s of exposure time for the blue spectral range between 3130 {\AA} and 5960 {\AA} and the red spectral
range between  4600 {\AA} and 9000 {\AA}, respectively. 

The spectra were extracted from the flat-fielded two-dimensional image of each exposure as described in \citet{Klein2010,Klein2011}.  The most challenging task was removing the
broad undulations in the continuum likely caused by variable vignetting \citep{Suzuki2003}.  Also, as described in \citet{Klein2010}, an additional re-normalizing processing step was applied to calibrate and remove second (diffraction) order flux contamination in the region 8200 - 9000 {\AA}.  Wavelength calibration was performed using the standard Th-Ar lamps.  Following \citet{Klein2010,Klein2011}, we used IRAF to normalize the spectra and combine echelle orders.

With our signal to noise ratio which varied between 30 and 45, lines with an equivalent width as weak as 20 m{\AA} or even a little less can be detected.  As a result,  well over 100 lines  from 11 
elements heavier than helium are seen; there are no unidentified lines.

As noted by the referee, archived {\it Hubble Space Telescope} data acquired with the {\it Cosmic Origins Spectrograph} were
acquired under program ID\# 11561 with B. Gaensicke as PI.   Although now available to the public, we have not included
these data in our formal analysis.   We did find that our model fit to the carbon and oxygen abundances derived from the Keck observations agree very well with the
ultraviolet observations.  However, because it is of great value to the  scientific community to have  independent  abundance studies based on different data
and model atmospheres, we have reported only analysis of the Keck observations.

\section{ELEMENTAL ABUNDANCES}
We first tried to estimate the atmospheric parameters by fitting the {\it Sloan Digital Sky Survey} (SDSS)
spectroscopic data with a grid appropriate for DB white dwarf stars.
However, given that many of the helium lines are contaminated by metal
absorption lines, we do not believe that the standard spectroscopic
technique can provide accurate atmospheric parameters, especially for the
surface gravity. We thus decided to instead obtain the effective
temperature by fitting the ugriz photometry and keeping $\log$ $g$ fixed at
8.0. We obtain $T_{eff}$ = 19,535 ${\pm}$ 700K. It is well known however that the
presence of metals at the photosphere of a white dwarf can have a
significant impact on the thermodynamic structure, leading to an
overestimation of the effective temperature \citep{Wegner1985,
Provencal2002, Dufour2005, Dufour2010}. We thus calculated a new
DB grid with heavy elements assuming abundances close to our final adopted
values. Fitting the ugriz data with this grid we obtain, as expected, a
lower effective temperature of 18,700K ${\pm}$ 700K, which we adopt for the rest of
our analysis. Figure 1 shows a synthetic spectrum with these parameters
over the SDSS data where most of the Si, Ca and Fe lines are nicely
reproduced, indicating that our assumptions were good and, most
importantly, that the thermodynamic structure used for the rest of our
analysis is more realistic. We also verified a posteriori that the
measured abundances of lines from different ions (MgI/MgII and FeI/FeII)
agreed very well (within 0.1 dex), indicating that the atmospheric
parameters we have assumed are not significantly different than their true
values.

\begin{figure}
 \plotone{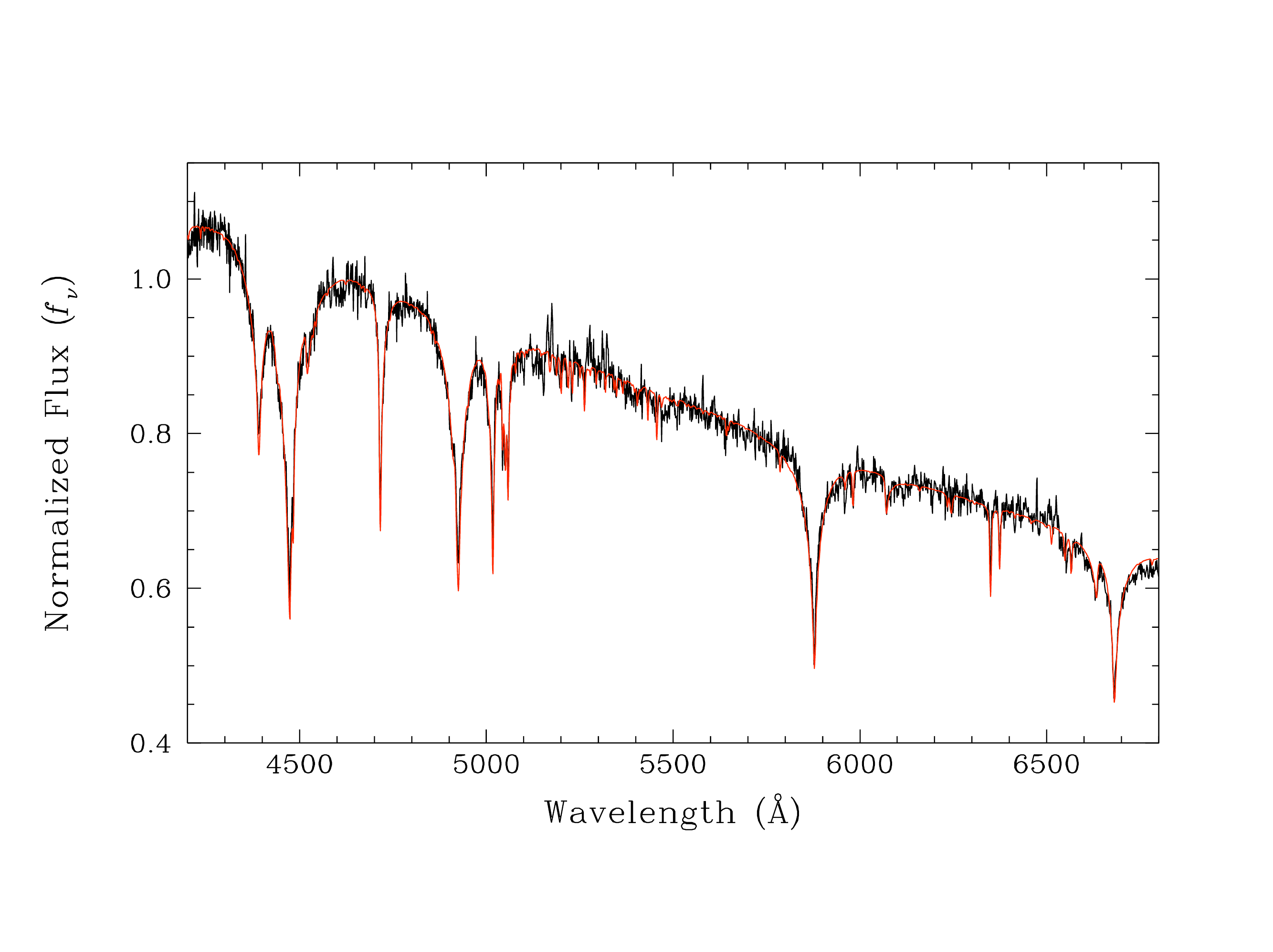}
\caption{SDSS optical spectrum of Ton 345.  Black denotes the data; the agreement of the He line profiles with the model denoted in red, supports  our inferred atmospheric parameters.  } 
\end{figure}

Using the thermodynamic structure of a model calculated with $T_{eff}$ =
18,700 K, $\log$ $g$ = 8.0, $\log$ H/He = -5.0 and the approximate amount of heavy
elements discussed above, we next analyze the Keck observations following
the procedure described in Dufour et al. (2012). The heavy element
abundances found in this way were then used to compute a new model
atmosphere (we keep $T_{eff}$ and $\log$ $g$ fixed to the values cited above). We
next repeat the fitting procedure using this new atmospheric structure and
find that the abundances remain practically unchanged, indicating that we
have converged to a final solution (see Table 1) and that no further
iteration is required. Most of the lines are well separated from each
other and therefore we can infer a value of an elemental abundance from an
individual line. The dispersion in the various measurements for a given
element can roughly be used as minimum abundance uncertainty, which is
typically around 0.1 dex.  For some elements, such as O, the dispersion of
the abundances determined from individual lines can be as small as 0.02
dex. However, given the uncertainties in continuum placement, atomic
parameters and the model atmosphere, we adopt a minimum error associated
with each abundance determination of 0.1 dex. There are also uncertainties
associated with the effective temperature and gravity, but, fortunately,
the relative abundances are insensitive to small variations in these
parameters (Klein et al. 2011). We show model fits to the spectral lines
listed in Table 2 in Figures 2-9.

As far as we know, Ton 345 is the only externally-polluted white dwarf to display optical carbon lines as seen in Figure 2.  The carbon to oxygen ratio in Ton 345 is a factor of 10 -100 greater than found in other heavily polluted white dwarfs \citep{Jura2014}.  The H${\alpha}$ line is at best only very marginally detected; we can only place an upper bound to the hydrogen abundance.  Some of the Si II lines are not well fit; previous studies also have found difficulties in simultaneously fitting all available silicon lines in the spectra of externally polluted white dwarfs \citep{Jura2012a,Gaensicke2012}.

\begin{figure}
 \plotone{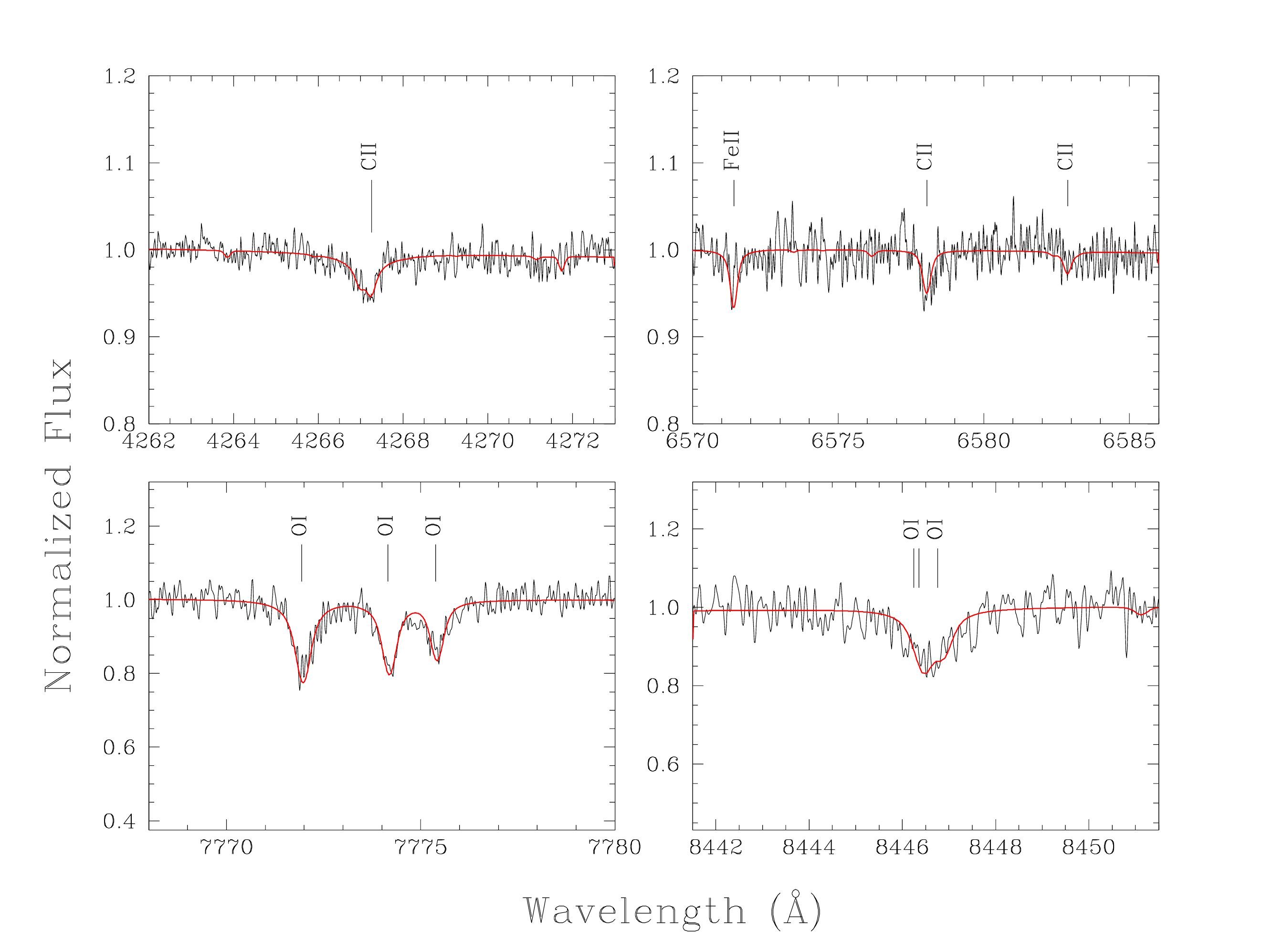}
\caption{Spectrum of Ton 345  with  lines of C, O,  and Fe.  Black denotes the data while red denotes the model with our inferred abundances.  The model is wavelength shifted to the photospheric frame  of the star; wavelengths are in air and the heliocentric frame of rest. } 
\end{figure}

\begin{figure}
 \plotone{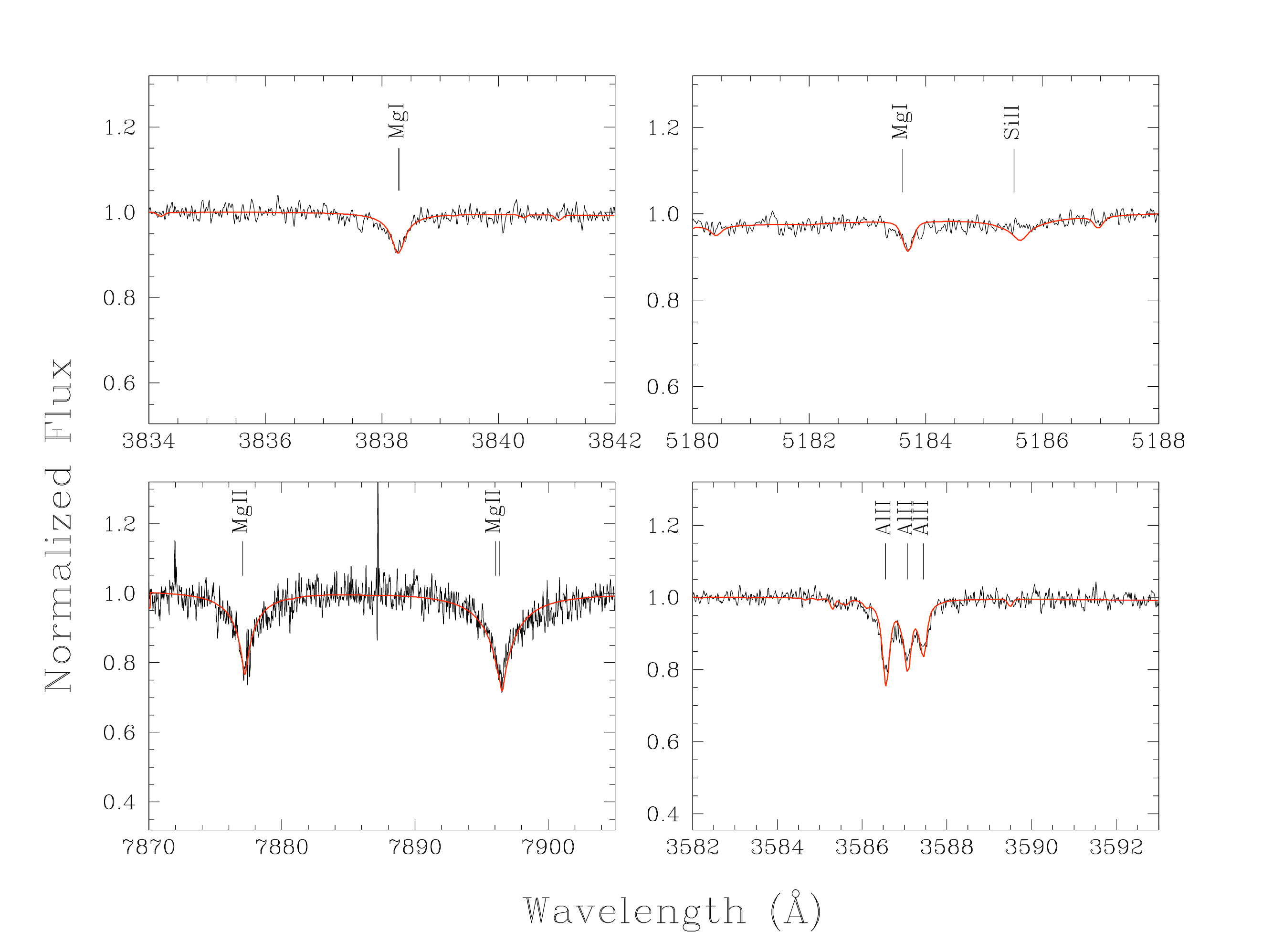}
\caption{The same as Figure 2  with lines of Mg, Al, and Si.} 
\end{figure}

\begin{figure}
 \plotone{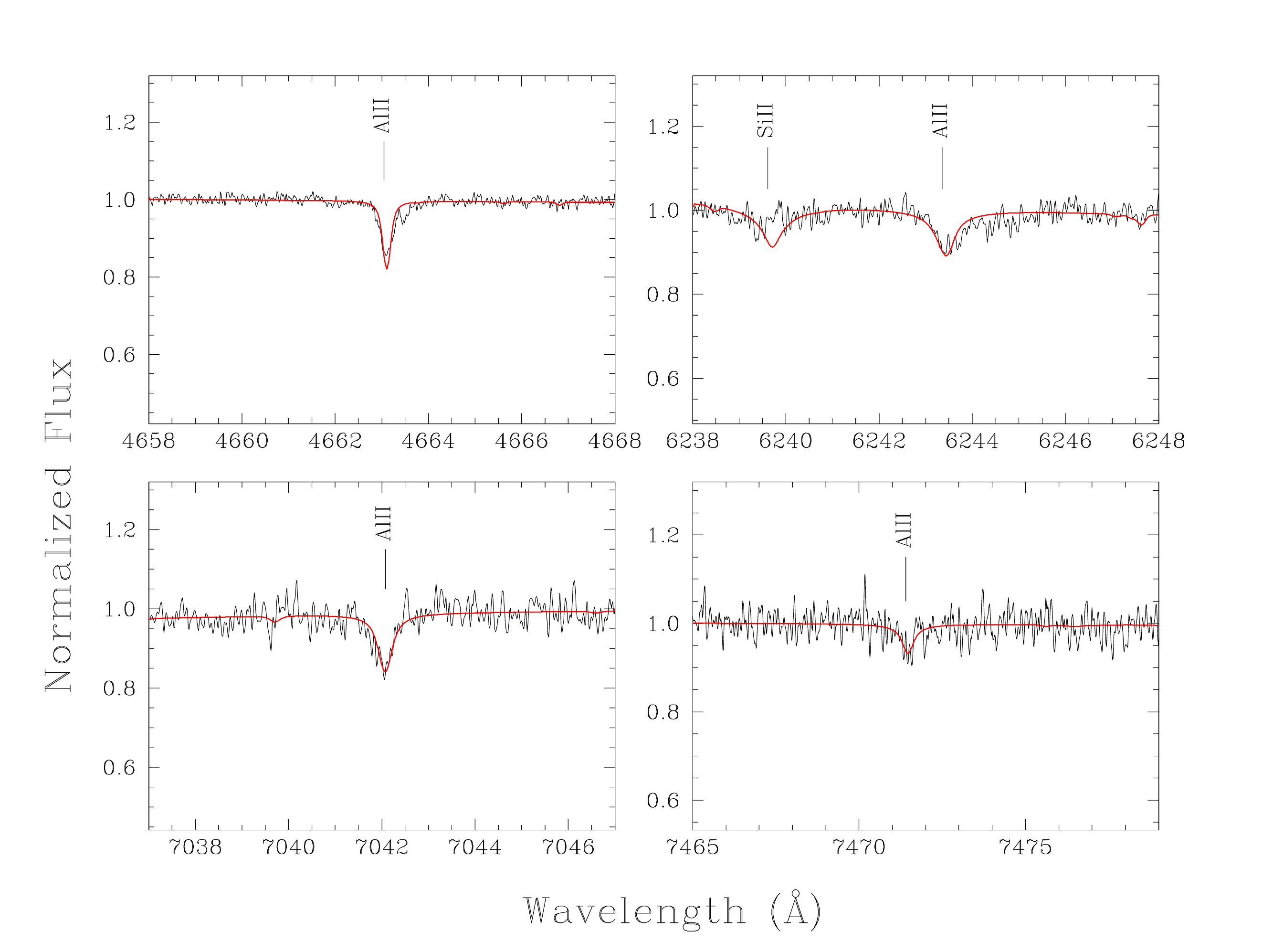}
\caption{The same as Figure 2 with lines of Al and Si.} 
\end{figure}

\begin{figure}
 \plotone{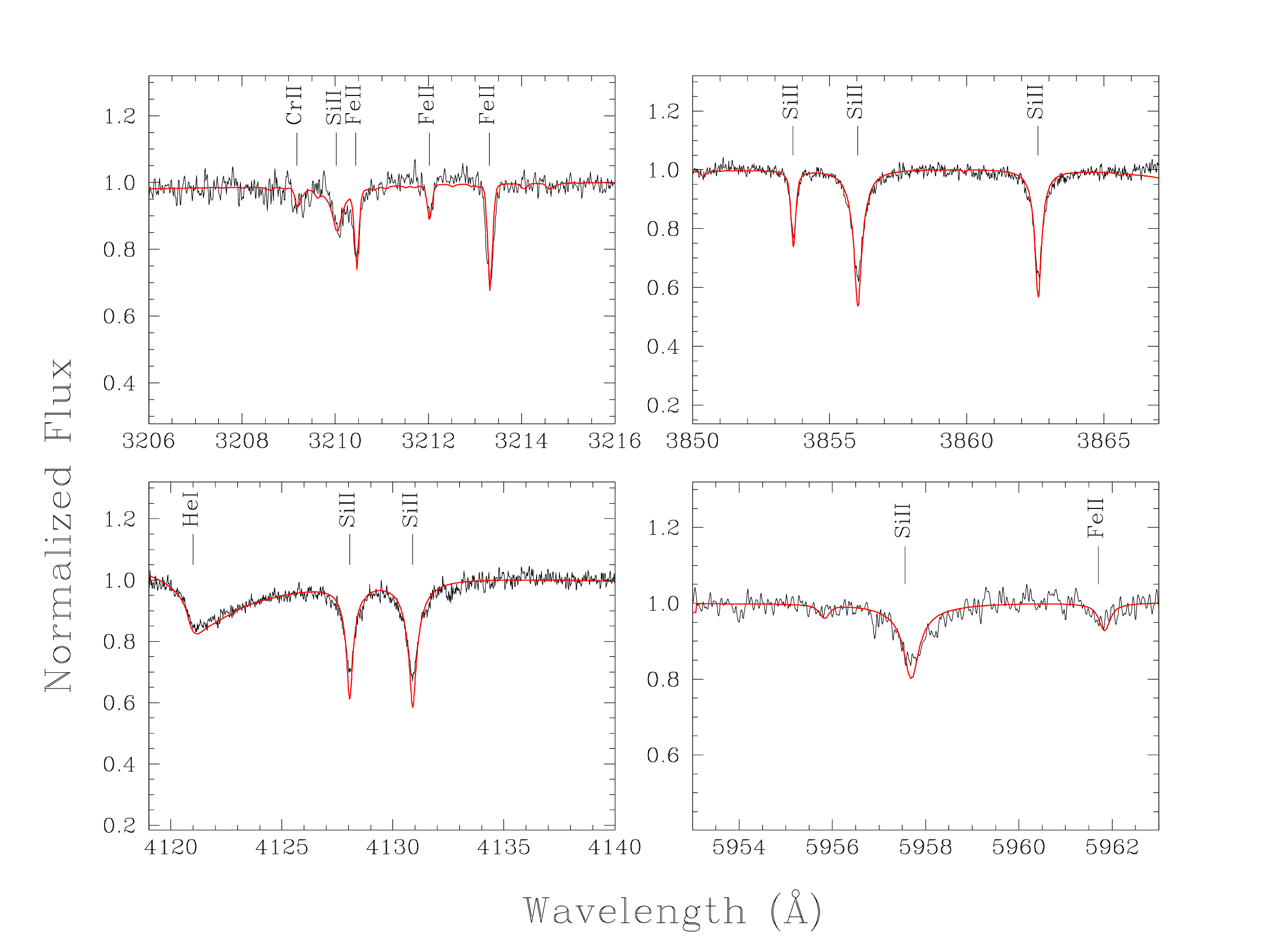}
\caption{The same as Figure 2 with lines of He, Si, Cr and Fe.} 
\end{figure}
\begin{figure}
 \plotone{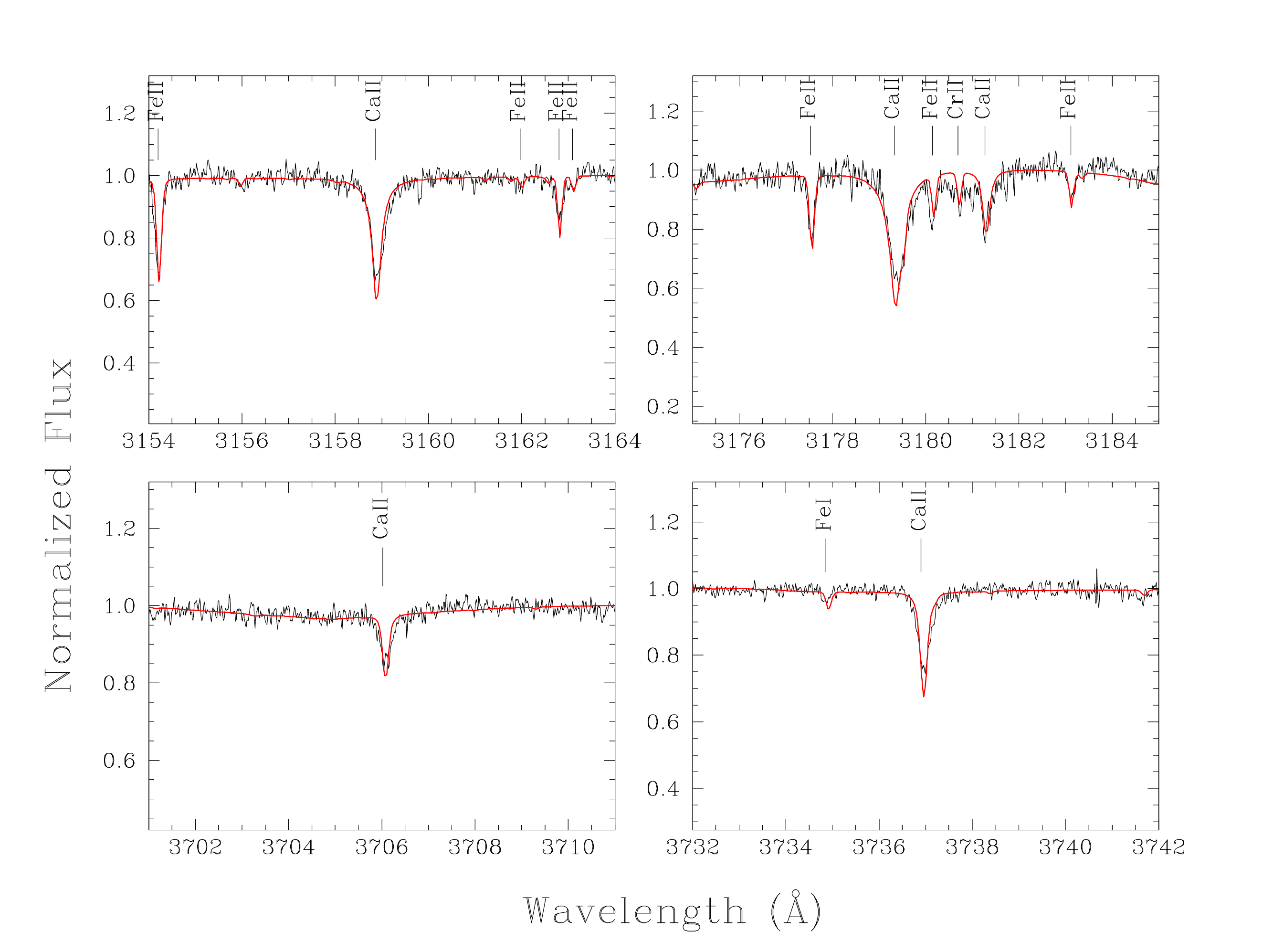}
\caption{The same as Figure 2 with  lines of Ca, Cr and Fe.} 
\end{figure}

\begin{figure}
 \plotone{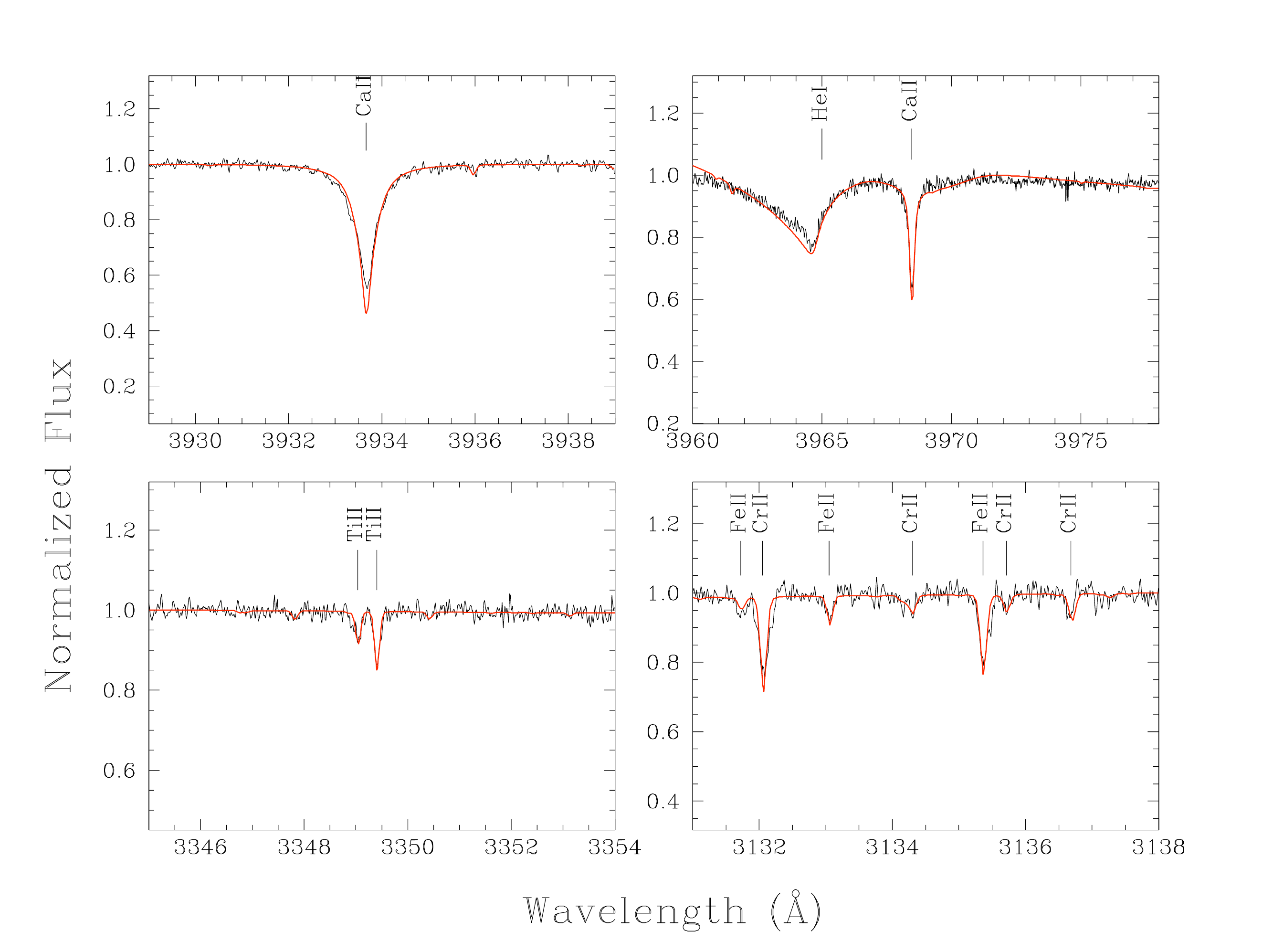}
\caption{The same as Figure 2 with lines of He, Ca, Ti, Cr and Fe.} 
\end{figure}
\begin{figure}
 \plotone{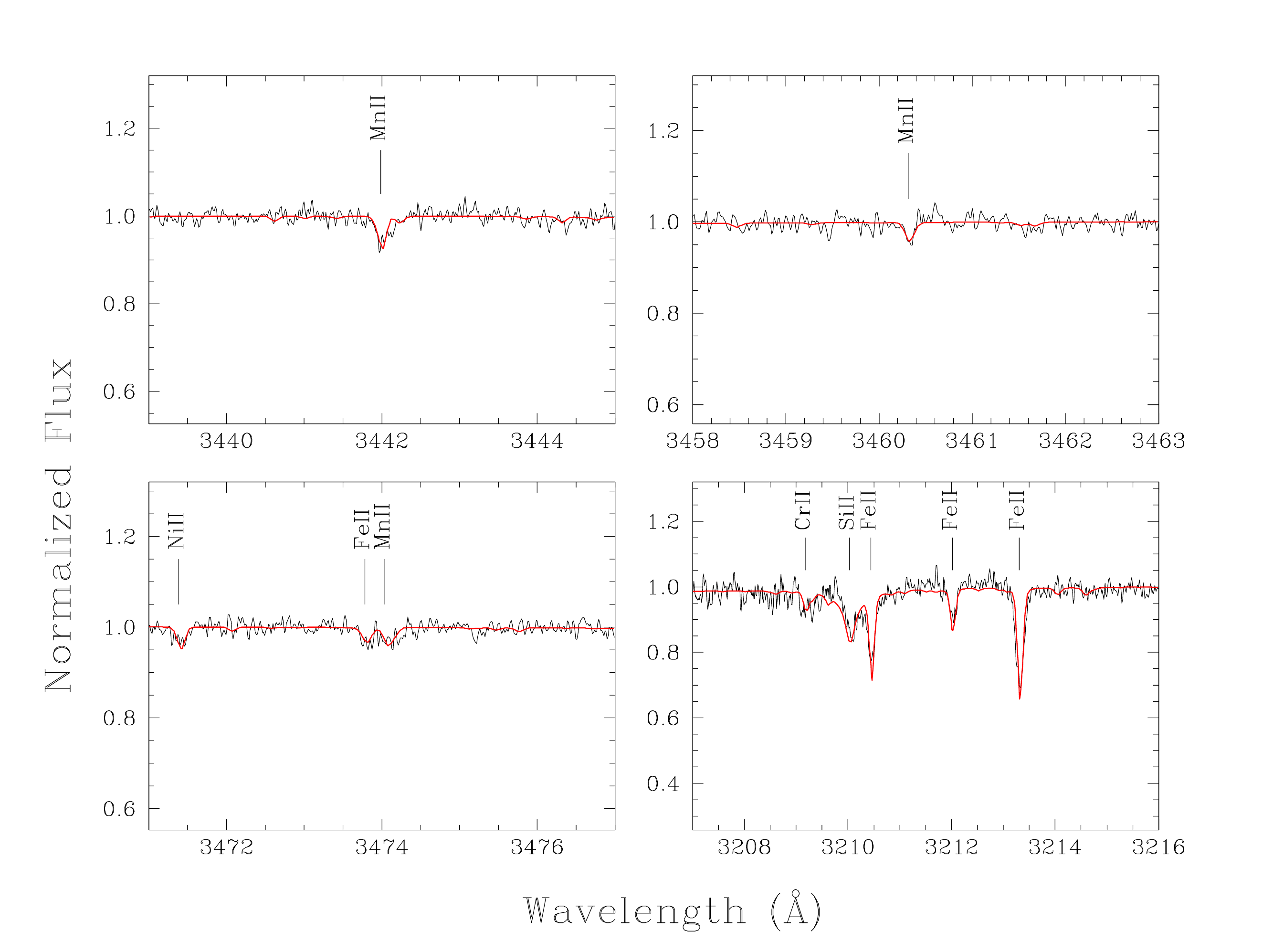}
\caption{The same as Figure 2 with  lines of Si, Mn, Cr, Fe, and Ni.} 
\end{figure}
\begin{figure}
 \plotone{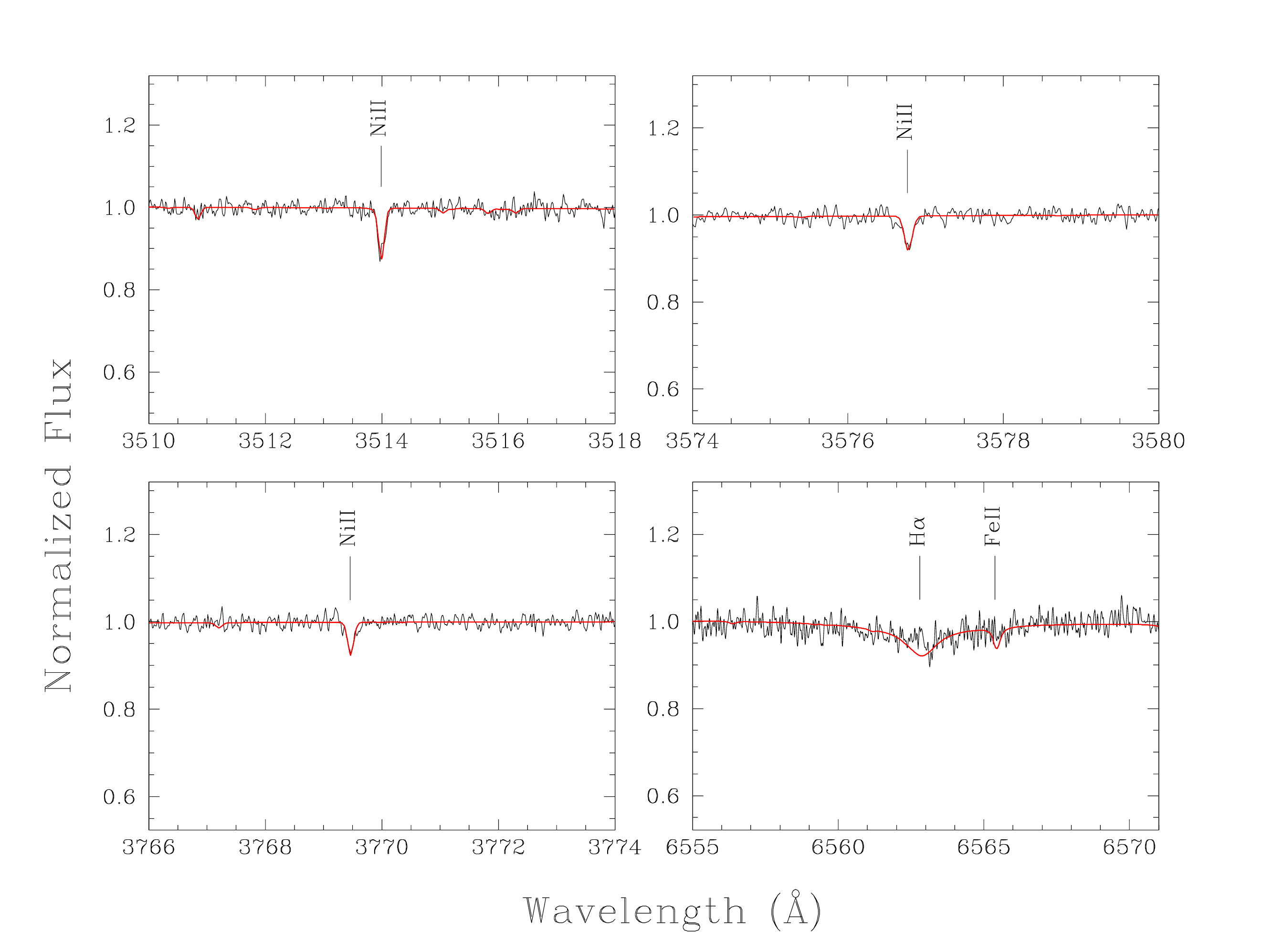}
\caption{The same as Figure 2 except for selected lines of H, Fe and Ni.  The fit to H${\alpha}$ represents our upper bound to the abundance of this element which is only very marginally detected.} 
\end{figure}
\begin{center}
Table 1 -- Abundances in Ton 345
\\
\begin{tabular}{rrrrr}
\hline
\hline
Element & [$\log$ $n$(Z)/$n$(He)] \\ 
\hline
H & ${\leq}$- 5.47  \\
C & -4.63 (0.19)\\
O & -4.58 (0.10) \\
Mg & -5.02 (0.10) \\
Al & -5.96 (0.10) \\
Si & -4.91 (0.12)  \\
Ca & -6.23 (0.10) \\
Ti & -7.74(0.10) \\
Cr & -6.91 (0.10)\\
Mn & -7.54 (0.10)\\
Fe & -5.07 (0.10)\\
Ni & -6.20 (0.10)  \\
\hline
\end{tabular}
\end{center}
In this Table, we follow astronomical convention and report abundances by number.  Key lines used in the abundance determinations include: H${\alpha}$; C II 4267 {\AA}, 6578 {\AA}; O I 7771 {\AA}, 7774 {\AA}, 7775 {\AA}; Mg I 3838 {\AA}, 5184 {\AA}; Mg II 7877 {\AA}, 7896 {\AA}; Al II  4663 {\AA}, 6243 {\AA}, 7042 {\AA}, 7471 {\AA}; Si II 3210 {\AA}, 3854 {\AA}, 3856 {\AA}, 3863 {\AA }, 4128 {\AA}, 4131 {\AA}, 5958 {\AA}, Ca II 3159 {\AA}, 3179 {\AA}, 3181 {\AA}, 3706 {\AA}, 3737 {\AA}, 3933 {\AA}, 3968 {\AA}; Ti II, Cr II, Fe II: multiple lines \citep{Klein2010}; Mn II 3442 {\AA}, 3460 {\AA}, 3474 {\AA}; Ni II 3514 {\AA}.
\section{THE ACCRETED PARENT BODY}
Because Ton 345 possesses a dust disk, it is likely that the pollution results from one large parent body with a well defined angular momentum vector; otherwise grains likely would be destroyed by mutual collisions \citep{Jura2008}.

While dredge-up might  enhance the carbon abundance in white dwarfs with effective temperatures near 25,000 K, this process is probably negligible for stars cooler than 20,000 K \citep{Koester2014b}.  We therefore proceed by assuming that all of 
the heavy elements in the atmosphere of Ton 345 are accreted from its circumstellar disk.  

Because different heavy elements settle at different rates, the abundances within the photosphere of an externally-polluted white dwarf do not necessarily directly reflect
the abundances in the parent body \citep{Koester2009}.  For Ton 345, the typical settling time\footnote{settling times are taken from http://www1.astrophysik.uni-kiel.de/~koester/astrophysics/} is 10$^{5}$ yr, comparable to estimates for  the lower bound of a typical dust disk lifetime  \citep{Girven2012}.   
Possibly, the outer convective zone might be in a steady state where the rate of
accretion is balanced by the rate of gravitational settling.  Alternatively, because it is both  observed and theoretically predicted that accretion rates onto externally-polluted white dwarfs can be variable on time scales much shorter
than 10$^{5}$ yr \citep{Metzger2012,Rafikov2012,Wilson2014,Xu2014b}, the abundances in the atmosphere of Ton 345 might reflect a recent burst of accretion.  

Here, we assume the ``instantaneous" approximation where the abundances in the photosphere equal the abundances in the parent body.  
If the system is in a steady state, the relative abundances of the lighter elements, C through Si, would be unchanged because their relative settling times differ by less than a factor of 1.1 from their mean value.  In contrast,  the relative abundances of the heavier elements such as Ca and Fe would
increase by a factor of two.  However, even though the fraction of the mass of the  parent body mass carried in these heavy elements would be larger,  our most important results -- that carbon
is unusually abundant and the material has little water, would be unaltered.  

There are two arguments that the parent body accreted onto Ton 345 was anhydrous.  If the matter accreted onto the white dwarf is carried within familiar minerals, then
Mg, Al, Si, and Ca are bonded to oxygen in the proportions matching the oxides MgO, Al$_{2}$O$_{3}$,  SiO$_{2}$ and CaO.  Iron may be found either as an oxide or in metallic form.  By this mineralogical argument, we find that all the oxygen is bound into minerals; none is left to form water.  Also, the accreted minor planet was low in water because there is relatively little hydrogen in the atmosphere of Ton 345; from Table 1 and that most of the oxygen was bound in oxides, we
compute that less than 10\% of the oxygen was in the form of H$_{2}$O.

We now consider the composition of the accreted parent body.  Using  the abundances listed in Table 1, we compute the  mass fraction of each element as provided in  Table 2.  
Because of its large carbon abundance, it is possible that the planetesimal accreted onto Ton 345 resembles
the most primitive  meteorites, the  CI chondrites.   However, such a  fit to our data is not very good, because CI
chondrites have relatively more oxygen and less carbon then we measure in Ton 345.  Among five well studied CI chondrites, the  carbon mass percentage
is 3.5\% with a dispersion of  0.48\% and a maximum value of 4.4\%  \citep{Lodders2003},  much less than inferred for the material accreted onto Ton 345.  Among the same five CI chondrites, the average oxygen  mass 
percentage is 46\%  with a dispersion of 5.8 \% and a minimum value of  41\%\citep {Lodders2003} is notably higher than our value of 23\% inferred for the minor planet accreted onto
Ton 345.

Because of its high carbon to oxygen abundance ratio,
 the most familiar solar system material that matches the composition seen in Ton 345 is anhydrous Interplanetary Dust Particles (IDPs), primitive matter whose average
mass fractions \citep{Thomas1993} also are given in Table 2.  
Approximately 50\% of IDPs are anhydrous \citep{Flynn2003}, and these are the ones we consider here.  The elemental mass fractions for  Ton 345's pollution  and of  anhydrous IDPs agree except for Ni.

For our model atmosphere, we compute that the He mass in the outer convection zone is 9.1 ${\times}$ 10$^{25}$ g.  Consequently, from the abundances given in  Table 1, the total mass of
the accreted parent body must have been at least 1.6 ${\times}$ 10$^{23}$ g, about 60\% of the mass of Vesta \citep{Russell2012}  If its density was 3 g cm$^{-3}$, then the parent body diameter was at least 470 km,  well within the range inferred for these parameters for  extrasolar planetesimals  accreted onto heavily polluted white dwarfs \citep{Jura2014}.  

\begin{center}
Table 2 -- Percentage of Total Mass
\\
\begin{tabular}{lrrrrr}
\hline
\hline
Element & Ton 345 & IDP \\
&\%  &\%) \\
\hline
C & 15 (5.6) &  12.5 (5.7) &  \\
O & 23 (4.0) &  32.9 (4.0) & \\
Mg & 13 (2.5)  & 10.7 (4.6)& \\
Al & 1.6 (0.36) & 1.3 (1.1) & \\
Si & 19 (4.2) & 14.6 (2.9) & \\
Ca &1.3 (0.29) & 0.9 (0.3) & \\ 
Cr & 0.35 (0.079) & 0.2 (0.1)  & \\
Mn & 0.086 (0.020) & 0.2 (0.3) &  \\
Fe & 26 (4.4) & 17.6 (6.3) &   \\
Ni & 2.0 (0.45) & 0.7 (0.4) &  \\
\hline
\end{tabular}
\end{center}
Following cosmochemical convention, the abundances ratios are expressed by mass rather than by element number.  The values for Ton 345 are derived from Table 1.  Except for H and Ti whose abundances were not reported, the values for anhydrous  IDPs  in the third column are taken from Table 3 of \citet{Thomas1993} with their 1${\sigma}$ dispersion given in parentheses.  
\section{DISCUSSION}
Previous observations have shown extrasolar minor planets accreted by white dwarfs  typically have little water \citep{Jura2012b}, although  the parent body accreted onto GD 61 is an exception in being 
water-rich  \citep{Farihi2013}.  In contrast,
the object accreted onto Ton 345  is carbon-rich and water-poor indicating that the  carbon to water
content in extrasolar minor planets varies by more than a factor of 100.

A star's C/O ratio influences the carbon and water content of planetesimals that are formed in its protoplanetary nebulae \citep{Johnson2012}.  However, essentially
all main sequence stars near the sun have more oxygen than carbon \citep{Fortney2012, Nissen2013, Teske2014} and the observed range  in the carbon to oxygen ratio among externally polluted white dwarfs
must be largely a consequence of substantial  separation of these two elements during formation and evolution.

Heavily polluted white dwarfs with dust disks typically have low carbon abundances \citep{Jura2014}.  In contrast,  \citet{Koester2014b}
have found that some white dwarfs where they only report C and Si abundances with relatively modest levels of pollution have notably higher carbon to silicon abundance ratios
than do  systems where abundances of multiple elements have been measured.  One possible explanation for the  pollution with relatively high carbon abundance is that the parent bodies originate in the outer planetary system.  However, we do not
yet have a full understanding of the cosmochemical evolution of carbon in extrasolar planetary systems.

Because we have measured abundances of 11 heavy  elements in the atmosphere of Ton 345, we can  constrain many potential scenarios for the origin of the accreted minor planet accreted.   
\citet{Brown2012} has suggested that high-density KBOs with relatively little water are the consequence of the collisional erosion
of a differentiated parent body which had an ice shell and a rocky core.  Because differentiation is widespread among 
extrasolar planetesimals \citep{Jura2013}, some related model may pertain to the minor planet accreted onto
Ton 345.     We show in Figure 10 a comparison
between the mass fractions listed in Table 2 for Ton 345 and those for an anhydrous IDP, normalized to the abundances in CI chondrites.   The good agreement can be understood
as being the consequence of the evolution of a differentiated minor planet which initially had an iron core, a rocky mantle and an ice
exterior. At some later time,  all the ice, some of the mantle and none of the core was lost either by a collision or some other process such as sublimation of surface water ice  during the star's highly luminous red giant
phase before it became a white dwarf \citep{Jura2004} although buried water ice could be retained \citep{Jura2010}. Regardless of how this hypothetical differentiated planetesimal  lost its outer ice, ultimately, it would achieve a bulk composition resembling those solar system KBOs with relatively high density and little water.
\begin{figure}
 \plotone{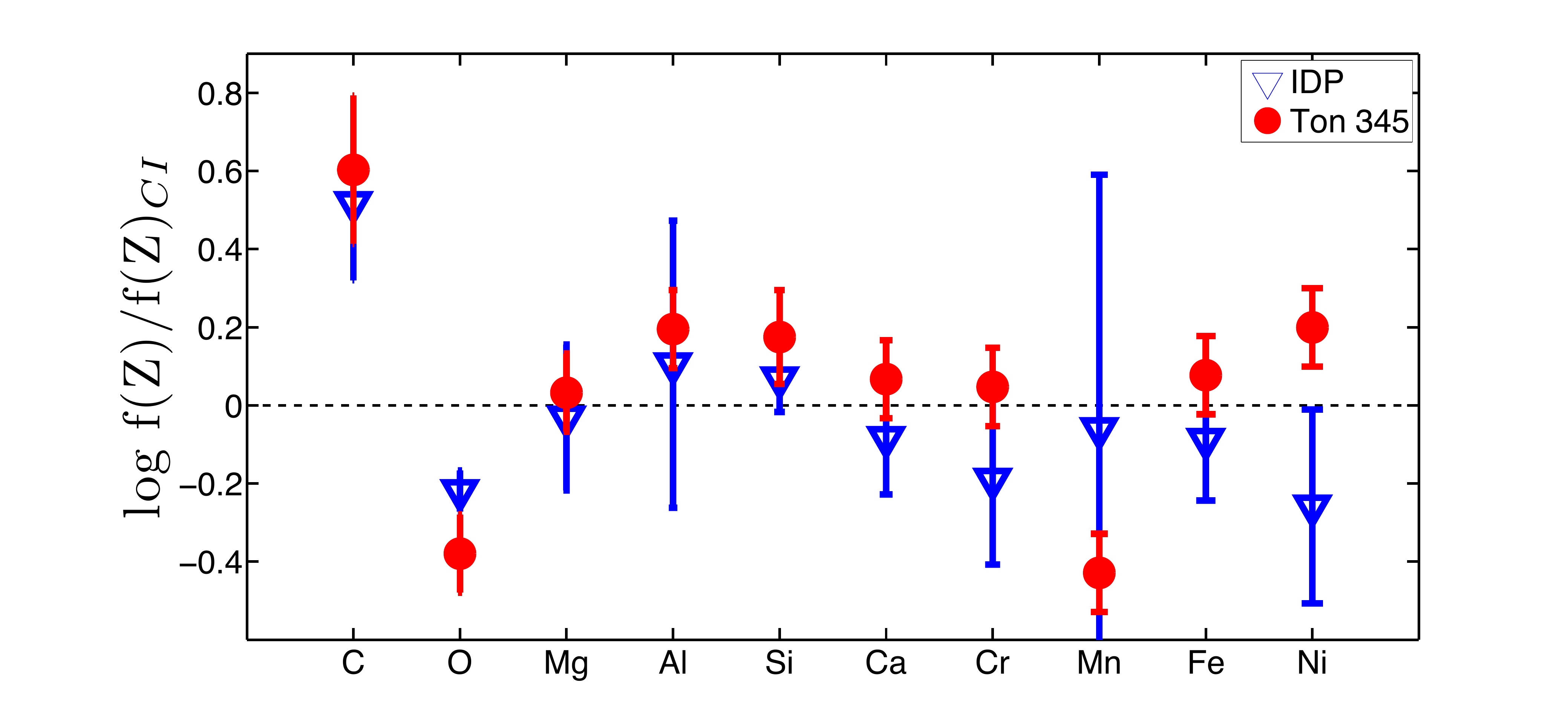}
\caption{Comparison of the mass percentages in Ton 345 with those in anhydrous IDPs scaled to the average values in CI chondrites.    We see that both the minor planet accreted onto Ton 345 and anhydrous IDPs are enriched in carbon and depleted in oxygen relative to CI chondrites.   } 
\end{figure}

\section{CONCLUSIONS}

We have obtained optical spectra of Ton 345 and measured the abundances of 11 elements heavier than helium.  We find that we are observing the disintegration of
a minor planet that  likely was carbon-rich and ice-poor; it appears to have been compositionally similar to
 a high density Kuiper Belt Object.

This work has been partly supported by the NSF. We thank M. Nabeshima for help with the data analysis.

 \bibliographystyle{apj}

\begin{thebibliography}{99}
\bibitem [{{Bergin} (2013)}] {Bergin2013} Bergin, E. A. 2013, to be published, XVII Special Courses at the National Observatory of Rio de Janeiro. AIP Conference Proceedings, astro-ph 1309.4729
\bibitem [{{Bonsor} {et~al.} (2011)}] {Bonsor2011} Bonsor, A., Mustill, A. J., \& Wyatt, M. C. 2010, \mnras, 414, 930
\bibitem [{{Brinkworth} {et~al.}(2012)}] {Brinkworth2012} Brinkworth, C. S., Gaensicke, B. T., Girven, J. M. et al. 2012, \apj, 750, 86
\bibitem [{{Brown} (2012)}] {Brown2012} Brown, M. E. 2012, Ann. Rev. Earth Plan. Sci., 40, 467
\bibitem [{{Debes} \& {Sigurdsson}(2002)}] {Debes2002} Debes, J. H., \& Sigurdsson, S. 2002, \apj, 572, 556
\bibitem [{{Dufour} {et~al.} (2005)}] {Dufour2005} Dufour, P., Bergeron, P., \& Fontaine, G. 2005, \apj, 627, 404
\bibitem [{{Dufour} {et~al.} (2010)}] {Dufour2010} Dufour, P., Kilic, M., Fontaine, G. et al. 2010, \apj, 719, 803
\bibitem [{{Dufour} {et~al.} (2012)}] {Dufour2012} Dufour, P., Kilic, M., Fontaine, G. et al. 2012, \apj, 749, 6
\bibitem [{{Farihi} {et~al.} (2010)}] {Farihi2010} Farihi, J., Jura, M., Lee, J.-E., \& Zuckerman, B. 2010, \apj, 714, 1386
\bibitem [{{Farihi} {et~al.} (2013)}] {Farihi2013} Farihi, J., Gaensicke, B. T., \& Koester, D. 2013, Science, 342, 218
\bibitem [{{Flynn} {et~al.} (2003)}] {Flynn2003} Flynn, G. J., Keller, L. P., Feser, M., Wirick, S., \& Jacobsen, C. 2003, Geochim. Cosmochim. Acta, 67, 479
\bibitem [{{Fortney} (2012)}] {Fortney2012} Fortney, J. 2012, \apj, 747, L27
\bibitem [{{Frewen} \& {Hansen} (2014)}] {Frewen2014} Frewen, S. F. N., \& Hansen, B. M. S. 2014, \mnras, 439, 2442
\bibitem [{{Gaensicke} {et~al.} (2008)}] {Gaensicke2008} Gaensicke, B. T., Koester, D., Marsh, T. R., Rebassa-Mansergas, A. \& Southworth, J. 2008, \mnras, 391, L103
\bibitem [{{Gaensicke} {et~al.} (2012)}] {Gaensicke2012} Gaensicke, B. T., Koester, D., Farihi, J. et al. 2012, \mnras, 424, 333
\bibitem [{{Girven} {et~al.} (2012)}] {Girven2012} Girven, J., Brinkworth, C. S., Farihi, J. et al. 2012, \apj, 749, 154
\bibitem [{{Henning} \& {Semenov} (2013)}] {Henning2013} Henning, T. \& Semenov, D. 2013, Chem. Rev., 113, 9016
\bibitem [{{Johnson} {et~al.} (2012)}] {Johnson2012} Johnson, T. V., Mousis O., Lunine, J., \& Madhusudhan, N. 2012, \apj,  757, 192
\bibitem [{{Jura}(2003)}] {Jura2003} Jura, M. 2003, \apj, 584, L91
\bibitem [{{Jura} (2004)}] {Jura2004} Jura, M. 2004, \apj, 603, 729
\bibitem [{{Jura} (2008)}] {Jura2008} Jura, M. 2008, \aj, 135, 1785
\bibitem [{{Jura} \& {Xu} (2010)}] {Jura2010} Jura, M. \& Xu, S. 2010, \aj, 140, 1129
\bibitem [{{Jura} \& {Xu} (2012)}] {Jura2012b} Jura, M. \& Xu, S. 2012, \aj, 143, 6
\bibitem [{{Jura} {et~al.} (2012)}] {Jura2012a} Jura, M., Xu, S., Klein, B., Koester, D., \& Zuckerman, B. 2012, \apj, 750, 69
\bibitem [{{Jura} {et~al.} (2013)}] {Jura2013} Jura, M., Xu, S., \& Young, E. D. 2013, \apj, 775, L41
\bibitem [{{Jura} \& {Young}(2014)}] {Jura2014} Jura, M. \& Young, E. D. 2014, Ann. Rev. Earth Plan. Sci., 42, 45
\bibitem [{{Klein} {et~al.} (2010)}] {Klein2010} Klein, B., Jura, M., Koester, D., Zuckerman, B., \& Melis, C. 2010, \apj, 709, 950
\bibitem [{{Klein} {et~al.} (2011)}] {Klein2011} Klein, B., Jura, M., Koester, D., \& Zuckerman, B. 2011, \apj, 741, 64
\bibitem [{{Koester} (2009)}]  {Koester2009} Koester, D. 2009, \aap, 498, 517
\bibitem [{{Koester} {et~al.} (2014a)}] {Koester2014a} Koester, D., Gaensicke, B., \& Farihi, J. 2014a, \aap, 566, 34
\bibitem [{{Koester} {et~al.} (2014b)}] {Koester2014b} Koester, D., Provencal, J., \& Gaensicke, B. T. 2014b, \aap, 568, 118
\bibitem [{{Lacerda} \& {Jewitt} (2007)}] {Lacerda2007} Lacerda, P. \& Jewitt, D. C. 2007, \aj, 133, 1393
\bibitem [{{Lee} {et~al.} (2010)}] {Lee2010} Lee, J.-E., Bergin, E. A., \& Nomura, H. 2010, \apj, 710, L21
\bibitem [{{Lockwood} {et~al.} (2014)}] {Lockwood2014} Lockwood, A. C., Brown, M. E., \& Stansberry, J. 2014, Earth Moon Planets, 111, 127
\bibitem [{{Lodders} (2003)}] {Lodders2003} Lodders, K. 2003,  \apj, 591, 1220
\bibitem [{{Melis} {et~al.} (2010)}] {Melis2010} Melis, C., Jura, M., Albert, L., Klein, B., \& Zuckerman, B. 2010, \apj, 722, 1078
\bibitem [{{Metzger} {et~al.}(2012)}] {Metzger2012} Metzger, B. D., Rafikov, R. R., \& Bochkarev, K. V. 2012, \mnras, 423, 505
\bibitem [{{Nissen} (2013)}] {Nissen2013} Nissen, P. E. 2013, \aap, 552, 73
\bibitem [{{Provencal} {et~al.} (2002)}] {Provencal2002} Provencal, J. L., Shipman, H. L., Koester, D., Wesemael, F., \& Bergeron, P. 2002, \apj, 568, 324
\bibitem [{{Rafikov} \& {Garmilla} (2012)}] {Rafikov2012} Rafikov, R. R., \& Garmilla, J. A. 2012, \apj, 760, 123
\bibitem [{{Russell} {et~al.} (2012)}] {Russell2012} Russell, C. T., Raymond, C. A., Coradini, A. et al. 2012, Science, 336, 684
\bibitem [{{Sicardy} {et~al.} (2011)}] {Sicardy2011} Sicardy, B., Assafin, M., Jehin, E. et al. 2011, Nature, 478, 493
\bibitem [{{Suzuki} {et~al.} (2003)}] {Suzuki2003} Suzuki, N., Tytler, D., Kirkman, D., O'Meara, J. M., \& Lubin, D. 2003, \pasp, 115, 1050
\bibitem [{{Teske} {et~al.} (2014)}] {Teske2014} Teske, J. K., Cunha, K., Smith, V. V. Schuler, S. C., \& Griffith, C. A. 2014, \apj, 788, 93
\bibitem [{{Thomas} {et~al.} (1993)}] {Thomas1993} Thomas, K. L., Blanford, G. E., Keller, L. P., Klock, W. \& McKay, D. S. 1993, Geochim. Cosmochim. Acta, 57, 1551
\bibitem [{{Veras} \& {Wyatt} (2012)}] {Veras2012} Veras, D., \& Wyatt, M. C. 2012, \mnras, 421, 2969
\bibitem [{{Vogt} {et~al.} (1994)}] {Vogt1994} Vogt, S. S., Allen, S. L., Bigelow, B. C. et al. 1994, SPIE, 2198, 362
\bibitem [{{Wegner} \& {Koester} (1985)}] {Wegner1985} Wegner, G., \& Koester, D. 1985, \apj, 288, 746
\bibitem [{{Wilson} {et~al.} (2014)}] {Wilson2014} Wilson, D. J.,Gaensciek, B., Koester, D. et al. 2014, \mnras, in press
\bibitem [{{Xu} {et~al.} (2013)}] {Xu2013} Xu, S., Jura, M., Klein, B., Koester, D., \& Zuckerman, B. 2013, \apj, 766, 132
\bibitem [{{Xu} \& {Jura} (2014)}] {Xu2014b} Xu, S., \& Jura, M. 2014, \apj, 792, L39
\bibitem [{{Xu} {et~al.} (2014)}] {Xu2014a} Xu, S., Jura, M., Koester, D., Klein, B., \& Zuckerman, B. 2014, \apj, 783, 79
 \end{thebibliography}

\end{CJK}
\end{document}